\documentclass[12pt]{article}
\usepackage{amsmath,amssymb,amsfonts,amscd,cancel,graphicx,hyperref,longtable}
\newcommand{\ba}{\begin{array}}
\newcommand{\ea}{\end{array}}
\newcommand{\be}{\begin{equation}}
\newcommand{\ee}{\end{equation}}

\begin{document}


\thispagestyle{empty}
\renewcommand{\thefootnote}{\fnsymbol{footnote}}
\setcounter{footnote}{1}

\vspace*{-0.5cm}

\centerline{\Large\bf Unified Model of Fermion Masses with} 
\vspace*{2mm}
\centerline{\Large\bf Wilson Line Flavor Symmetry Breaking}

\vspace*{18mm}

\centerline{\large\bf
Gerhart Seidl\footnote{E-mail: \texttt{seidl@physik.uni-wuerzburg.de}}
}

\vspace*{5mm}
\begin{center}
{\em Institut f\"ur Theoretische Physik und Astrophysik}\\
{\em Universit\"at W\"urzburg, D 97074 W\"urzburg, Germany} 
\end{center}

\vspace*{20mm}

\centerline{\bf Abstract}
\vspace*{2mm}
We present a supersymmetric $SU(5)$ GUT model with a discrete non-Abelian
flavor symmetry that is broken by Wilson lines. The model is formulated
in 4+3 dimensions compactified on a manifold $S^3/Z_n$. Symmetry
breaking by Wilson lines is topological and allows to realize the
necessary flavor symmetry breaking without a vacuum alignment mechanism. The
model predicts the hierarchical pattern of charged fermion masses and quark mixing angles. Small normal hierarchical neutrino masses are
generated by the type-I seesaw mechanism. The non-Abelian flavor
symmetry predicts to leading order exact maximal atmospheric mixing
while the solar angle emerges from a variant of quark-lepton complementarity. As a
consequence, the resulting leptonic mixing matrix is in excellent agreement with current data and could be tested in future neutrino oscillation experiments.

\renewcommand{\thefootnote}{\arabic{footnote}}
\setcounter{footnote}{0}

\newpage

\section{Introduction}
\label{sec:introduction}
One of the main motivations for considering physics beyond the standard
model (SM) is to understand the  observed pattern of fermion
masses. Among the key features of the fermion sector that a theory
more fundamental than the SM should account for are the observed large
hierarchies of the quark masses and mixing angles in the
Cabibbo-Kobayashi-Maskawa (CKM) matrix $V_\text{CKM}$ \cite{CKM}. Moreover, one would like to know how this can
be related to the leptonic Pontecorvo-Maki-Nakagawa-Sakata
(PMNS) \cite{PMNS} mixing matrix $U_{\text{PMNS}}$, which, as neutrino
oscillation experiments have told us, contains two large mixing
angles. A theory of fermion masses should provide a rationale for
these observations, especially in the light of quark-lepton
unification in grand unified theories (GUTs) \cite{SU5,Pati:1974yy}.

Current neutrino oscillation data can be well approximated
 \cite{Schwetz:2008er} by a tribimaximal PMNS matrix
 \cite{Harrison:1999cf}, which suggests an interpretation in terms of
 a non-Abelian flavor symmetry. In fact, there have been a number of
 attempts to arrive at tribimaximal lepton mixing by making use of
 non-Abelian discrete flavor symmetries such as the tetrahedral group $A_4$ \cite{A4}, the double
(or binary) tetrahedral group $T'$ \cite{T'}, or $\Delta(27)$
\cite{delta27} (for reviews see \cite{Ma:2007ia}). A general
difficulty of these models is to achieve the correct
symmetry breaking in order to predict, in addition to the fermion
 mixing angles, also the observed fermion mass hierarchy. The
 difficulty is that non-Abelian flavor
 symmetries relate the Yukawa couplings of different generations,
 which generally produces fermion masses of the same order and no
 hierarchy between them. This problem becomes particularly severe when trying a unified description of quark and
lepton masses (see, e.g., \cite{Feruglio:2007uu}) in GUTs \cite{discreteGUTs}.

In four-dimensional (4D) models, fermion mass hierarchies can be
generated from non-Abelian flavor symmetry breaking by employing a
vacuum alignment mechanism. The implementation of a proper 4D vacuum
alignment mechanism, however, appears in practice often
complicated in unified models. This difficulty can be avoided in an
extra-dimensional setting. In fact, the localization properties
of fields in certain five-dimensional (5D) limits of multi-throat geometries
\cite{Cacciapaglia:2006tg} (see also \cite{Kaplan:2001ga} and
\cite{Agashe:2007jb}) allow to realize the necessary non-Abelian
flavor symmetry breaking without the need for a vacuum alignment
mechanism \cite{Plentinger:2008nv} (for related studies of orbifolds see \cite{Kobayashi:2008ih}).

In this paper, we make use of the Hosotani mechanism \cite{Hosotani:1983xw} or symmetry breaking by Wilson lines \cite{Candelas:1985en,Witten:1985xc,superstringtheory,Ibanez:1986tp} to break a
discrete non-Abelian flavor symmetry of a supersymmetric $SU(5)$
GUT toy model in 4+3 dimensions. Originally, Wilson line breaking has been used for GUT
breaking and as a solution to the doublet-triplet splitting problem
\cite{Candelas:1985en,Witten:1985xc,Ibanez:1986tp}. By applying Wilson line
breaking to a discrete non-Abelian flavor symmetry \cite{Hall:2001tn} that has been gauged \cite{Krauss:1988zc},
we can reproduce the observed fermion mass hierarchies along with an
exact leading order prediction for leptonic mixing, without the need
for a vacuum alignment mechanism. Main features of the fermion sector will thus become topological in origin.

The paper is organized as follows. In Sec.~\ref{sec:geometry}, we
introduce the geometry of the higher-dimensional model along with the
particle content and localization of fields. Next, in
Sec.~\ref{sec:flavorsymmetry}, we present our non-Abelian flavor symmetry together with predictions from higher-dimension operators. In
Sec.~\ref{sec:wilsonlines}, we discuss the breaking of the non-Abelian
flavor symmetry by Wilson lines. Our results for the masses and
mixings of quarks and leptons are given in Sec.~\ref{sec:masses+mixings}. Finally, in Sec.~\ref{sec:conclusions}, we present our summary and conclusions.

\section{Geometry and Fields}\label{sec:geometry}
Let us consider a (4+3)-dimensional space-time compactified to four dimensions on a manifold $Q=Q_0/Z_n$, where
$Q_0=S^3$ is a three-sphere. In choosing the topology of our setup,
the geography of fields, and to implement a successful Wilson line breaking
mechanism, we follow closely the example in \cite{Witten:2001bf} (see
also \cite{Barbieri:1994jq,Ross:2004mi}). The three-sphere can be described by two complex
numbers $z_1$ and $z_2$ with $|z_1|^2+|z_2|^2=1$. After modding out with
respect to a group $L\simeq Z_n$ that acts freely on $z_{1,2}$ by
\begin{equation}\label{eq:L}
 L:z_i\rightarrow e^{2\pi i/n}z_i,\quad i=1,2,
\end{equation}
we arrive at the physical space $Q=Q_0/L=S^3/Z_n$. In addition, we assume a global symmetry
$F\simeq Z_n$ that acts by
\begin{equation}\label{eq:F}
 F\::\: z_1\rightarrow z_1,\,z_2\rightarrow e^{2\pi i/n}z_2.
\end{equation}
Obviously, $Q$ is left invariant by $F$. The symmetry $F$ has as fixed points two
circles $S_1$ and $S_2$, where $S_1$ is given by $|z_1|=1,z_2=0$ and
$S_2$ is given by $z_1=0,|z_2|=1$. Note that $S_1$ is invariant under
$F$, while $S_2$ is left invariant by $F$ up to the transformation (\ref{eq:L}).
As the symmetry group of the model we choose $SU(5)\times G_F$, where
$G_F$ is a non-Abelian discrete flavor symmetry group (for seminal
work on 5D orbifold GUTs see \cite{Kawamura:2000ev}). The chiral $SU(5)$ matter
superfields are ${\bf 10}_i$ and $\overline{\bf 5}_i$, where $i=1,2,3$ is the generation index, and ${\bf 5}^H$ and
$\overline{\bf 5}^H$ are the two Higgs superfields, giving
mass to the up- and down-type fields, respectively. To generate small
neutrino masses via the seesaw mechanism \cite{typeIseesaw,typeIIseesaw}, we introduce three right-handed
neutrino superfields ${\bf 1}_i$, which are $SU(5)$ singlets. Finally,
we assume flavon Higgs superfields $\phi_1$ and $\phi_2$, which
transform under the non-Abelian group $G_F$ as doublets $\phi_i=(\phi_i^a,\phi_i^b)^T$
($i=1,2$) but are singlets under $SU(5)$. The flavor symmetry group $G_F$
will be specified later. We suppose that all
three generations of supermultiplets ${\bf 10}_i$ and $\overline{\bf 5}_i$,
as well as ${\bf 5}^H$ and $\overline{\bf 5}^H$, are localized on the circle $S_1$. In contrast to this, the three right-handed neutrino
superfields ${\bf 1}_i$ and the flavon superfields $\phi_1$ and
$\phi_2$ are assumed to be localized on the circle $S_2$. The localization
of matter superfields is summarized in Tab.~\ref{tab:localization}.

\begin{table}
\begin{center}
\begin{tabular}{|c||c|c|}
\hline
\text{Field}&${\bf 10}_i,\overline{\bf 5}_i,{\bf 5}^H,\overline{\bf 5}^H$&
${\bf 1}_i,\phi_{1,2}$\\
\hline
\text{Location}&$S_1$&$S_2$\\
\hline
\end{tabular}
\caption{Localization of matter superfields on the circles $S_1$ and $S_2$.}\label{tab:localization}
\end{center}
\end{table}

\section{Non-Abelian Flavor Symmetry}\label{sec:flavorsymmetry}
According to the classification theorem of finite simple groups
(for an outline see \cite{classification}) each finite simple group is
isomorphic to one of the following four types: (1) A group of prime
order. (2) An alternating group. (3) A group of Lie type. (4) One of
26 sporadic groups. From the list of finite simple groups, we can
construct new finite groups through group extensions. For example,
$A_4,T',$ and $\Delta(27)$, which have been recently used as discrete flavor
symmetries, can be written in terms of semi-direct products as
\begin{equation}\label{eq:semidirect}
A_4\simeq Z_3\ltimes(Z_2\times Z_2),\quad T'\simeq Z_2\ltimes
Q,\quad\Delta(27)\simeq Z_3\ltimes(Z_3\times Z_3),
\end{equation}
where $Q$ is the quaternion group of order eight. Generally, let $X$ and $H$ be two discrete groups and $\varphi:X\rightarrow\text{Aut}\,H$
a homomorphism of $X$ into the automorphism group $\text{Aut}\,H$ of $H$, which maps $x$ onto $x^\varphi$. If we define on the Cartesian product
$K=\{(x,h)|\,x\in X,\,h\in H\}$ a multiplication law by
\begin{equation}
(x_1,h_1)(x_2,h_2)=(x_1x_2,h_1^{x_2}h_2)\quad(x_i\in X,h_i\in H),
\end{equation}
where $h_1^{x_2}=(x_2^\varphi)^{-1}h_1 x_2^\varphi$, then $K$ is
called a semi-direct product
$K=X\ltimes H$ (with respect to $\varphi$) \cite{Kurzweil}. When $\varphi$
is the trivial homomorphism, $X\ltimes H$ becomes the direct product $X\times H$.

In our model, we construct a non-Abelian flavor symmetry group $G_F$ by taking the semi-direct product
\begin{equation}\label{eq:GF}
G_F=\tilde{G}\ltimes G
\end{equation}
of two discrete groups $\tilde{G}$ and $G$. For $\tilde{G}$ we choose
\begin{equation}\label{eq:Gtilde}
\tilde{G}=Z_3\times Z_8\times Z_9,
\end{equation}
where the three generations of matter fields carry under $\tilde{G}$ the charges
\begin{eqnarray}\label{eq:fermioncharges}
&{\bf 10}_1\sim(1,1,6),\;\;{\bf 10}_2\sim(0,3,1),\;\;{\bf
10}_3\sim(0,0,0),&\nonumber\\
&\overline{\bf 5}_1\sim(1,4,2),\;\;\overline{\bf
5}_2\sim(0,7,0),\;\;\overline{\bf 5}_3\sim(0,0,1),&\\
&{\bf 1}_1\sim(2,0,6),\;\;{\bf 1}_2\sim(2,6,0),\;\;{\bf
    1}_3\sim(2,0,6),&\nonumber
\end{eqnarray}
while the components of the flavon fields have the charge assignment
\begin{eqnarray}
&\phi_1^a\sim(1,1,3),\;\;\phi_1^b\sim(1,0,2),&\nonumber\\
&\phi_2^a\sim(0,6,8),\;\;\phi_2^b\sim(0,5,7).&
\end{eqnarray}
The Higgs fields ${\bf 5}^H$ and $\overline{\bf 5}^H$ are singlets
under $G_F$. Note that the group $\tilde{G}$ with its charge assignment to the $SU(5)$ matter
fields and right-handed neutrinos in (\ref{eq:fermioncharges}) is as in
\cite{Plentinger:2008nv}. The group $G$, on the other hand, is a direct product
\begin{equation}
G=G_1\times G_2\times G_3
\end{equation}
of discrete groups $G_i$. Let us combine $\overline{\bf 5}_2,\overline{\bf 5}_3,\phi_i^a,$ and $\phi_i^b$,
into the doublets $\overline{\bf 5}_d=(\overline{\bf
  5}_2,\overline{\bf 5}_3)^T$ and $\phi_i=(\phi_i^a,\phi_i^b)^T$. We suppose that the doublets $\overline{\bf 5}_d$ and $\phi_i$ transform under $G_1$ as
\begin{equation}\label{eq:G1}
 G_1:\overline{\bf 5}_d\rightarrow P\overline{\bf
   5}_d,\,\phi_i\rightarrow P\phi_i,
\end{equation}
where $i=1,2$, and $P=\left(\begin{matrix}0&1\\1&0\end{matrix}\right)$ is a $2\times 2$
permutation matrix. Since $G_1$ does not commute with $\tilde{G}$,
the total flavor symmetry group $G_F$ is non-Abelian and we have in
(\ref{eq:GF}) a semi-direct product with non-trivial homomorphism
$\tilde{G}\rightarrow\text{Aut}\,G$.

The groups $G_2$ and $G_3$ are isomorphic to $Z_p$ and $Z_q$, where
$p$ and $q$ are even, and act on the fields as
\begin{equation}\label{eq:G2G3}
G_2\simeq Z_p:{\bf 10}_2\rightarrow -{\bf
  10}_2,\,\,\phi_1\rightarrow-\phi_1,\qquad
G_3\simeq Z_q:{\bf 1}_1\rightarrow -{\bf 1}_1,\,{\bf 1}_3\rightarrow -{\bf 1}_3,\,
\phi_2\rightarrow -\phi_2,
\end{equation}
i.e.~${\bf 10}_2$ and $\phi_1$ carry the charge $p/2$ under $Z_p$ and ${\bf
  1}_1,{\bf 1}_3$, and $\phi_2$, carry the charge $q/2$ under $Z_q$. 
  
Let us now comment on the choice of the symmetry groups $\tilde{G}$ and $G$. We wish to stress that the group $\tilde{G}$ and its charge assignment is far from being unique. In fact, in \cite{Plentinger:2008nv}, we have studied $\tilde{G}$ as just one example out of several hundred possibilities that can equally well yield realistic quark and lepton masses and mixings. The group (\ref{eq:Gtilde}) with the charge assignment in (\ref{eq:fermioncharges}) has been found in an automated scan based on product groups of cyclic symmetries assuming principles of quark-lepton complementarity \cite{qlc}. The advantage of cyclic product groups is that they allow to reproduce the observed fermion mass and mixing parameters in a comparatively simple way. Alternatively, in \cite{Plentinger:2008up}, we have  also analyzed the possibility to obtain the lepton masses and mixings from products of continuous symmetries and found far less examples, which is a simple consequence of the fact that there are more ways to saturate the charges in the discrete case. Choosing a non-Abelian symmetry for $\tilde{G}$ in such a scan would, of course, be even more challenging because this would require to account also for a vacuum alignment of scalar fields, which is, in contrast, not necessary for an Abelian $\tilde{G}$. For the scan, we have only allowed the simplest (singly charged) flavon representations and all Yukawa couplings should be close to one. The scan starts with the lowest rank and searches then for groups with increasing rank. The smallest flavor group found in this way is  $Z_7\times Z_9$ \cite{Plentinger:2008up}. Out of the sample of possibilities, we are led to the particular example in (\ref{eq:Gtilde}) and (\ref{eq:fermioncharges}) by requiring that the structure of the fermion mass matrices be as hierarchical as possible and thus highly determined by the quantum numbers only. It is not easy to see why simpler groups than that would fail, but relaxing the conditions on the solutions mentioned above could very well increase the number of viable models with groups of smaller rank. However, some of the striking features of  the observed fermion masses are readily seen from (\ref{eq:fermioncharges}). For instance, since ${\bf 10}_3$ is a total singlet under $G_F$, we will have a large top Yukawa coupling. Also, notice that the hierarchical structure of the charged fermion masses and CKM angles is reflected in (\ref{eq:fermioncharges}) by the rough increase of the charges of ${\bf 10}_i$ and $\overline{\bf 5}_i$ when going from the heavier to the lighter generations.

Moreover, in searching for $\tilde{G}$, we have been looking for an example that realizes a so-called lopsided form of the down quark mass matrix which allows, in addition, an exchange symmetry such as $G_1$ in (\ref{eq:G1}) between the 2nd and 3rd generation. The exchange symmetry $G_1$ acts in the lepton sector as a $\mu-\tau$ symmetry and is, thus, mainly responsible for explaining the near maximal atmospheric lepton mixing. In the quark sector, however, $G_1$ leads only to an unobservable large mixing among the right-handed quarks. The fact that $G_1$ does not commute with $\tilde{G}$, thus giving a non-Abelian $G_F$, is a key feature of the symmetries ensuring that the atmospheric mixing angle is to leading order exactly maximal by virtue of $G_F$. The symmetries in
  (\ref{eq:G2G3}), on the other hand, commute with $\tilde{G}$. They can be viewed as auxiliary symmetries accounting for some detailed properties of the Yukawa interactions by letting $\phi_1$ and $\phi_2$ couple predominantly to ${\bf 10}_2$ and ${\bf 1}_{1,3}$, respectively.

To break the direct product $\tilde{G}\times G_2\times G_3$, we assume the simplest implementation of
the Froggatt-Nielsen mechanism \cite{Froggatt:1978nt}: Every single
$Z_n$ group in this
product comes with a pair of $SU(5)$ singlet flavon superfields $f$
and $\overline{f}$, which are singly charged as $+1$ and $-1$ under the $Z_n$
symmetry. Under all other symmetries, the fields $f$ and
$\overline{f}$ transform trivially.

In the 4D low energy effective theory, we arrive at the Yukawa couplings
\begin{equation}\label{eq:4DLagrangian}
\mathcal{L}=\int
d^2\theta{\big [}Y^{u}_{ij}{\bf 10}_i{\bf 10}_j{\bf
5}^H+Y^{d}_{ij}{\bf 10}_i\overline{\bf 5}_j\overline{\bf 5}^H
+Y^\nu_{ij}\overline{\bf 5}_i{\bf 1}_j{\bf 5}^H
+M_{B-L}Y^R_{ij}{\bf 1}_i{\bf 1}_j+\text{h.c.}{\big]},
\end{equation}
where $Y^x_{ij}$, with $x=u,d,\nu,R$, are the dimensionless Yukawa
coupling matrices and $M_{B-L}\simeq 10^{14}\,\text{GeV}$ is the $B-L$
breaking scale generating small neutrino masses via the type-I seesaw
mechanism \cite{typeIseesaw}. Consider first the limit of vanishing vacuum
expectation values (VEVs)
$\langle\phi_1\rangle=\langle\phi_2\rangle=0$. In this limit, the
Froggatt-Nielsen mechanism generates for large $p$ and $q$ in
(\ref{eq:G2G3}) Yukawa coupling textures $Y^x_{ij}\rightarrow\tilde{Y}^x_{ij}$ of the forms
\begin{equation}\label{eq:quarktextures}
\tilde{Y}^u_{ij}\sim\left(
\begin{array}{ccc}
\epsilon^6 & 0 & \epsilon^5\\
0 & \epsilon^4 & 0\\
\epsilon^5 & 0 & 1
\end{array}
\right),\;
\tilde{Y}^d_{ij}\sim\epsilon\left(
\begin{array}{ccc}
\epsilon^4 & \epsilon^3 & \epsilon^3\\
0 & 0 & 0\\
\epsilon^6 & 1 & 1
\end{array}
\right),
\end{equation}
\begin{equation}\label{eq:neutrinotextures}
\tilde{Y}^\nu_{ij}\sim\epsilon^3\left(
\begin{array}{ccc}
0 & \epsilon & 0\\
0 & \epsilon & 0\\
0&\epsilon&0
\end{array}
\right),\;
\tilde{Y}^R_{ij}\sim\epsilon^4\left(
\begin{array}{ccc}
1& 0 & 1\\
0 & \epsilon & 0\\
1 & 0 & 1
\end{array}
\right),
\end{equation}
where we have neglected $\mathcal{O}(1)$ coefficients and assumed, for simplicity, that the symmetry breaking
parameters of the different $Z_n$ groups are roughly of the same
size and of the order the Cabibbo angle $\epsilon\simeq
\theta_\text{C}\simeq 0.2$. A more realistic description of quark and
lepton masses may be achieved by using different expansion factors,
say, for the up- and down-type sectors.

The group $G_1$ establishes in
(\ref{eq:quarktextures}) to leading order the exact relation
\begin{equation}\label{eq:tilderelations}
 \tilde{Y}^d_{32}=\tilde{Y}^d_{33}.
\end{equation}
The symmetries $G_2$ and $G_3$, on the other hand, are responsible for
suppressing the 2nd row in $\tilde{Y}^d_{ij}$ and the 1st and 3rd
column of $\tilde{Y}^\nu_{ij}$. We place ${\bf 10}_1$ and ${\bf 1}_2$
on points of $S_1$ (${\bf 10}_1$) and $S_2$ (${\bf 1}_2$), where the permutation symmetry $G_1$ is locally
broken. Consequently, we can have
$\tilde{Y}^d_{12}/\tilde{Y}^d_{13}=\mathcal{O}(1)$ and
$\tilde{Y}^\nu_{22}/\tilde{Y}^\nu_{32}=\mathcal{O}(1)$ without the
need for these ratios to equal exactly one.

In (\ref{eq:quarktextures}) and (\ref{eq:neutrinotextures}), we have
also written in front of the charged fermion and neutrino Yukawa coupling
matrices the overall suppression factors that are produced by
$\tilde{G}$. Observe also that only the top Yukawa
coupling is not suppressed by the flavor symmetry.

\section{Flavor Symmetry Breaking by Wilson Lines}\label{sec:wilsonlines}
Consider a gauge group $G$ with gauge field $A$ on a manifold $Q$.
For a loop $\gamma$ on $Q$, the Wilson line is defined as
\begin{equation}\label{eq:Wilsonline}
U_\gamma=\mathcal{P}\,\text{exp}\Big{(}i\oint_\gamma A\cdot dx\Big{)},
\end{equation}
where $\mathcal{P}$ denotes path ordering. The Wilson line, or holonomy, thus maps $\gamma$ into $G$. Moreover, for vanishing field strengths (flat connections), application of Stoke's theorem shows that $U_\gamma$, when defined with respect to a fixed base point, is invariant under continuous deformation of $\gamma$. Under a gauge transformation by $U(x_0)\in G$ at a  point $x_0\in\gamma$, the Wilson line transforms as $U_\gamma\rightarrow U(x_0)U_\gamma U^{-1}(x_0)$.
For an Abelian gauge group, $U_\gamma$ is therefore gauge invariant but for non-Abelian $G$ it transforms non-trivially under the gauge group.  In the low-energy theory, the Wilson lines appear then as effective composite fields resembling Higgs fields that transform in the adjoint representation \cite{superstringtheory}.

For non-simply connected $Q$, there is the possibility of non-contractible loops $\gamma$.  If $\gamma$ is non-contractible, the Wilson line $U_\gamma\neq 1$ can, in general, not be set to unity by a gauge transformation, although the field strength may vanish locally, i.e. on $Q$. In this case, for $U_\gamma\neq 1$, the gauge group is broken to the subgroup that commutes with $U_\gamma$, which is known as the Hosotani
mechanism \cite{Hosotani:1983xw}, symmetry breaking by Wilson lines
\cite{Candelas:1985en,Witten:1985xc,superstringtheory}, or symmetry breaking by background gauge
fields \cite{Ibanez:1986tp}.  The mapping of a loop $\gamma$ into $G$ in (\ref{eq:Wilsonline}) defines a homomorphism $\pi_1(Q)\rightarrow G$ from the fundamental group $\pi_1(Q)$ (which describes the product, i.e. combination, of successive loops based at the same point) into the gauge group. The gauge-inequivalent ground-state configurations are then given by the set of all these homomorphisms modulo $G$ \cite{Hall:2001tn}.

Let us now see what happens for non-contractible $\gamma$ when we try to gauge away the gauge field $A$ as much as possible. If we set by a gauge transformation $A=0$ on all of $\gamma$, we will single out one point $x_0\in\gamma$ where, upon parallel transport along the loop, we have to perform a gauge transformation by $U_\gamma$ whenever crossing $x_0$. The group element $U_\gamma$ appears then as a ``transition function'' at this point \cite{Hall:2001tn}. This leads us to the description of the allowed states on a quotient manifold $Q=Q_0/L$, where $Q_0$ is a manifold and $L$ a freely acting discrete group (since $L$ acts freely, $Q$ is still a manifold). In this case, the quotient manifold is multiply connected with non-trivial fundamental group $\pi_1(Q)=L$. A possible ground state configuration is therefore characterized by a mapping $g\rightarrow U_g$, where $g\in L$ and $U_g\in G$. This mapping is a
homomorphism from the fundamental group $L$ into a discrete subgroup $\overline{L}$ of $G$ with $U_g\in \overline{L}\subset G$. Consider now on $Q$ a field $\psi(x)$, where $x\in Q$ and we have set $A=0$ everywhere by a gauge transformation. Since $g(x)=x$, application of $g$ on the coordinates $x$ is similar to the parallel transport around a closed loop involving the transition function described above and thus requires a gauge transformation by $U_g$. This means that $\psi$ has to satisfy the relation $\psi(g(x))=U_g\psi(x)$ \cite{Witten:1985xc}. In other words, after switching on the Wilson lines, i.e. for $U_g\neq 1$, the only allowed modes on $Q$ are those which are
$L+\overline{L}$ singlet states. This is another way of saying that
for any physical state the application of a discrete group transformation $g\in L$ must be accompanied by a symmetry transformation $\overline{g}\in\overline{L}$.

Symmetry breaking by Wilson lines can also be applied to discrete groups \cite{Hall:2001tn} by promoting them to discrete gauge groups in the sense of \cite{Krauss:1988zc}. For illustration, let us discuss a simple example in five dimensions where the fifth dimension has been compactified on a circle $S^1$ with radius $R$ and 5D coordinate $y\in[0,2\pi R]$. As the discrete gauge symmetry we take the dihedral group $D_4\simeq Z_2\ltimes Z_4$ of order eight, which is the symmetry group of the square. The group $D_4$ is non-Abelian and has one two-dimensional and four one-dimensional irreducible representations. It is generated by the two elements
\begin{equation}
g_1=\left(\begin{matrix}
0 & 1\\
1 & 0
\end{matrix}\right),\qquad
g_2=
\left(\begin{matrix}
1 &0\\
0 & -1
\end{matrix}\right).
\end{equation}
We gauge $D_4$ by embedding it into an $SU(2)$ gauge symmetry. In terms of the $SU(2)$ generators, the elements      $g_1$ and $g_2$ can be expressed as
\begin{equation}\label{eq:SU(2)}
g_{1,2}=\frac{1}{i}\text{exp}\left(i\vec{\theta}_{1,2}\frac{\vec{\sigma}}{2}\right)\in SU(2),
\end{equation}
where $\vec{\theta}_1=(\pi,0,0)^T$, $\vec{\theta}_2=(0,0,\pi)^T$, $\vec{\sigma}=(\sigma_1,\sigma_2,\sigma_3)^T$, and $\sigma_i$ ($i=1,2,3$) are the usual Pauli matrices. We assume that the $SU(2)$ gauge symmetry be broken to $D_4$ by the VEV of some suitable scalar field, thereby giving masses to all three gauge bosons of $SU(2)$. Additional contributions to the gauge boson masses then come from the Wilson line symmetry breaking of $D_4$. To see this, we take the fifth component of the $SU(2)$ gauge field along the 2nd isospin direction, i.e.
\begin{equation}
A_5(x^\mu,y)=A_5^2(x^\mu,y)\frac{\sigma_2}{2}=A_5^2(x^\mu,y)\frac{1}{2}\left(\begin{matrix}
0 &-i\\
i & 0
\end{matrix}\right).
\end{equation}
We also set $A^2_5(x^\mu,y)\equiv A^2_5(x^\mu)$ constant on the circle by applying a gauge transformation, i.e. we go to ``almost axial gauge''. The Wilson line is then
\begin{equation}
 U_\gamma=\text{exp}\left(i\oint A^2_5\frac{\sigma_2}{2}dy\right)=\left(\begin{matrix}
\text{cos}\,\pi  A^2_5 R& \text{sin}\,\pi  A^2_5R\\
-\text{sin}\,\pi  A^2_5 R & \text{cos}\,\pi  A^2_5R
\end{matrix}\right),
\end{equation}
where we have used the abbreviation $A^2_5=A^2_5(x^\mu)$. In the effective theory, the corresponding Wilson line operator therefore commutes with the 5D gauge field in the 2nd isospin direction $A^2_5(x^\mu)\sigma_2/2$ but does not commute with the gauge fields $A^i_5(x^\mu)\sigma_i/2$ for $i=1,3$. Therefore, from (\ref{eq:SU(2)}), the Wilson line $U_\gamma$ does not commute with the discrete group generators $g_1$ and $g_2$ of $D_4$ either. It commutes, however, with four elements that generate a $Z_4$ subgroup of $D_4$. The Wilson line $U_\gamma$ therefore breaks $D_4\rightarrow Z_4$.

To study the effect of the Wilson line symmetry breaking more explicitly, we expand the 4D components of the $SU(2)$ gauge field as $A_\mu(x_\mu,y)=A_\mu^{(0)}(x_\mu)+\sum_{n=1}^\infty(A_\mu^{(n)}(x_\mu)e^{iny/R}+\text{h.c.})$ ($\mu=0,1,2,3$), where  the ${A_\mu^{(n)}}(x_\mu)$ ($n>1$) are complex and we have used the notation $A^{(n)}_\mu(x_\mu)=A^{(n)a}_\mu(x_\mu)\sigma^a/2$. After integrating out the extra dimension, the 4D low-energy effective action for the zero modes $A_\mu^{(0)}(x_\mu)$ is in our gauge $S_\text{eff}=2\pi R\,\text{Tr}\int d^4x (-\frac{1}{2}F^{(0)}_{\mu\nu}F^{(0)\mu\nu}+(D_\mu A_5)^2)$, where $F^{(0)}_{\mu\nu}=\partial_\mu A^{(0)}_\nu-\partial_\nu A^{(0)}_\mu-ig[A^{(0)}_\mu,A^{(0)}_\nu]$ is the field strength, $D_\mu A_5=\partial_\mu A_5-ig[A^{(0)}_\mu,A_5]$ the covariant derivative, and $g$ the 4D  $SU(2)$ gauge coupling. Since $A_5$ commutes with $A^{(0)2}_\mu$, we observe that the covariant derivative generates for a nonzero VEV $\langle A_5\rangle\neq 0$ only gauge boson masses for the zero modes $A^{(0)1}_\mu$ and $A^{(0)3}_\mu$, whereas $A^{(0)2}_\mu$ does not receive a mass from Wilson line symmetry breaking. This shows that only the $SU(2)$ gauge bosons which are associated with the discrete group elements $g_1$ and $g_2$ obtain extra masses from Wilson line symmetry breaking.

Let us next investigate for our 5D example the generation of non-local Wilson line type operators by integrating out heavy fermions following \cite{Csaki:2002ur}. Different from the discussion so far, these operators shall now be associated with open paths (as opposed to closed loops) connecting two points in the extra dimensional space. For this purpose, we introduce a  5D fermion $\Psi$ with action
\begin{eqnarray}\label{eq:S5DPsi}
\mathcal{S}_\Psi&=&
\int d^4x\int_0^{2\pi R}dy\,\Big\{\overline{\Psi}(iD_\mu\gamma^\mu-m)\Psi-\overline{\Psi}\gamma_5\partial_5\Psi+ig\overline{\Psi}A_5\gamma_5\Psi\nonumber\\
&&\qquad+
(\delta(y_1-y)C_1\overline{\Psi}\xi+\delta(y_2-y)C_2\overline{\Psi}\chi+\text{h.c.})
\Big\},
\end{eqnarray}
where $m$ is the mass of the fermion, $D_\mu=\partial_\mu-igA_\mu$ the covariant derivative, $\gamma^\mu=\sigma^1\otimes\sigma^\mu$ ($\sigma^0=-\mathbf{1}_2$), and $\gamma^5=\text{diag}(-\mathbf{1}_2,\mathbf{1}_2)$. We have assumed two 4D fermions $\xi$ and $\chi$ located at $y_1$ and $y_2$ with dimensionful couplings $C_1$ and $C_2$ ($C_{1,2}=[m]^{1/2}$) to $\Psi$, respectively. Again, we assume a gauge where $A_5$ is constant along the extra dimension. The fermion $\Psi$ is $2\pi$ periodic in $y$ and can be expanded as $\Psi(x_\mu,y)=\sum_{n=-\infty}^{\infty}\Psi^{(n)}(x_\mu)e^{iny/R}$ (in a gauge where $A_5$ vanishes everywhere, $\Psi$ would be $2\pi$ periodic only up to a transition function \cite{Hall:2001tn}). From (\ref{eq:S5DPsi}), we thus obtain the 4D action
  \begin{eqnarray}\label{eq:S4DPsi}
\mathcal{S}_\Psi&=&
2\pi R\int d^4x \sum_{n=-\infty}^{\infty}\Big\{\overline{\Psi}^{(n)}(i\gamma^\mu\partial_{\mu}-m-i\frac{n}{R}\gamma_5+igA_5\gamma_5)\Psi^{(n)}\nonumber\\
&&\qquad+
(C_1e^{-iny_1/R}\overline{\Psi}^{(n)}\xi+C_2e^{-iny_2/R}\overline{\Psi}^{(n)}\chi+\text{h.c.})
\Big\},
\end{eqnarray}
where terms $\sim A_\mu$ have been neglected. After integrating out the KK fermions $\psi^{(n)}$, we arrive at a 4D mass term between $\xi$ and $\chi$ of the form
\begin{equation}\label{eq:Yukawa}
\mathcal{L}_Y=2 \pi R C_1^\dagger C_2\sum_{n=-\infty}^{n=\infty}
\frac{2m\cdot e^{in(y_1-y_2)/R}}{m^2 + (\frac{n}{R}-gA_5)^2}
\overline{\xi}\chi+\text{h.c.}
\end{equation}
In the limit where $|y_1-y_2|/R$ is small and $mR$ or $g A_5R$ are large, the sum in (\ref{eq:Yukawa}) can be approximated by an integral. For $y_1-y_2$ positive (negative), we pick up the pole $RgA_5\pm im$ in the upper (lower) half-plane resulting in the Wilson line type operator
\begin{equation}\label{eq:residuum}
\mathcal{L}_Y=C\cdot
\text{exp}\left(-m|y_1-y_2|\right)\cdot
\text{exp}\left(-ig\int_{y_1}^{y_2}A_5 dy\right)
\overline{\xi}\chi+\text{h.c.}+\dots,
\end{equation}
where $C=i(2\pi)^2 R C_1^\dagger C_2\cdot\text{sgn}(y_1-y_2)$.  The dots denote terms suppressed by $|y_1-y_2|/R$, $m$, or $gA_5$. Moreover, observe in (\ref{eq:residuum}) the additional suppression factor $\text{exp}(-m|y_1-y_2|)$ which can produce large hierarchies of Yukawa couplings, but we will not make use of this possibility here.

In the following, we will apply the above ideas and shall be concerned with symmetry breaking by Wilson lines for the discrete non-Abelian flavor symmetry group $G_F$ introduced in
Sec.~\ref{sec:flavorsymmetry}. It is thus assumed  that $G_F$ be gauged. On the quotient manifold $Q=Q_0/L=S^3/Z_n$, the inequivalent ground state configurations are given by the homomorphisms of  $\pi_1(Q)=L=Z_n$ onto a discrete
subgroup $\overline{L}$ of $G_F$. As explained above, in presence of the Wilson
lines, the only allowed modes on $Q$ are those which are
$L+\overline{L}$ singlet states. In our example, we will take
$\overline{L}\simeq Z_n$, which is generated by a $2\times 2$ holonomy matrix
$W\in G_F$ of the form
\begin{equation}
 W=\left(
\begin{matrix}
e^{i\alpha}&0\\
0&e^{i\beta}
\end{matrix}\right),
\end{equation}
where $\alpha$ and $\beta$, with $\alpha\neq \beta$, are suitable $n$th
roots of unity such that $W^n=1$. Since $W$ does not commute with the
permutation matrix $P$ in (\ref{eq:G1}), the Wilson lines break $G_1$ to nothing.

Consider next the global symmetry $F$ in (\ref{eq:F}). Since $F$ leaves
 $S_1$ invariant, we let $F$ act only trivially on the matter
 fields located there, i.e.~${\bf 10}_i,\overline{\bf 5}_i,{\bf 5}^H$,
 and $\overline{\bf 5}^H$, are all $F$ singlets. On the circle
$S_2$, the action of $F$ is similar to that of $L$ (modulo a global symmetry) and must
be accompanied by a flavor symmetry transformation $W$. In addition to
 this, we assume that under $F$, $\phi_1$ transforms as $e^{-i\alpha}$
 and $\phi_2$ transforms as $e^{-i\beta}$. The Wilson
lines therefore leave a symmetry
$F'$ unbroken which is at $S_1$ given by $F$ and at $S_2$ by $F$
times a symmetry transformation $W$. As in the solutions to the
 doublet-triplet splitting problem
 \cite{Candelas:1985en,Witten:1985xc,Ibanez:1986tp}, this allows to ``split'' the
Yukawa couplings of the component fields in the $G_F$ doublets
$\overline{\bf 5}_d$ and $\phi_i$, although they are
related by the non-Abelian flavor symmetry $G_F$ before Wilson line breaking.

Now, $\phi_i^a$ and $\phi_i^b$ carry different $F'$ charges. As a consequence, in the 4D low-energy
effective theory, the $F'$-preserving non-renormalizable effective Yukawa couplings to
the fields $\phi_i$ that are generated after Wilson line breaking are,
to leading order, given by
\begin{equation}\label{eq:F'preserving}
\mathcal{L}_{F'}=M_F^{-1}\int
d^2\theta{\big [}
(\overline{\bf 5}_2,\overline{\bf 5}_3)
\left(\begin{matrix}
1 & 0\\
\epsilon^2 &0
\end{matrix}\right)
\left(
\begin{matrix}
\phi_1^a\\
\phi_1^b
\end{matrix}
\right)
{\bf 10}_2
\overline{\bf 5}^H
+(\overline{\bf 5}_2,\overline{\bf 5}_3)
\left(\begin{matrix}
0 & \epsilon^2\\
0 & 1
\end{matrix}\right)
\left(
\begin{matrix}
\phi_2^a\\
\phi_2^b
\end{matrix}
\right)
({\bf 1}_1+{\bf 1}_3)
{\bf 5}^H
+\text{h.c.}{\big]},
\end{equation}
where $M_F$ is some large mass scale. Note that the symmetry group $G_2\times G_3$
requires the $F'$-conserving corrections to (\ref{eq:F'preserving}) to
be of the order $(\phi^k_i/M_F)^3$, where $k=a,b$. Observe also that the non-renormalizable Yukawa interactions in the effective Lagrangian in (\ref{eq:F'preserving}) emerge from Wilson line type operators associated with open paths connecting two points on distinct circles $S_1$ and $S_2$ similar to the example leading to (\ref{eq:residuum}).

The crucial point in (\ref{eq:F'preserving}) is that the unbroken symmetry $F'$ allows the couplings
${\bf 10}_2\overline{\bf 5}_2\phi_1^a\overline{\bf 5}^H$,
$\overline{\bf 5}_3\phi_2^b{\bf 1}_1{\bf 5}^H$, and $\overline{\bf
  5}_3\phi_2^b{\bf 1}_3{\bf 5}^H$, while forbidding any couplings to
the components $\phi^b_1$ and $\phi^a_2$.\footnote{The approximately conserved symmetry
  $\tilde{G}\times G_2\times G_3$, which is left unbroken by the Wilson
  lines, leads in (\ref{eq:F'preserving}) to a suppression of the
  nonzero off-diagonal terms by a relative factor $\epsilon^2$.} For almost
arbitrary, non-vanishing VEVs $\langle\phi_i^k\rangle \neq 0$ that are roughly of the same order,
this will produce at low energies the Yukawa coupling terms $\epsilon'{\bf 10}_2\overline{\bf 5}_2\overline{\bf 5}^H$,
$\epsilon'\overline{\bf 5}_3{\bf 1}_1{\bf 5}^H$, and $\epsilon'\overline{\bf
  5}_3{\bf 1}_3{\bf 5}^H$, where we have introduced the small
parameter $\epsilon'\simeq \langle\phi^k_i\rangle/M_F$. The
corrections to these couplings are suppressed by relative factors
$\epsilon^2$ (from the off-diagonal elements) and ${\epsilon'}^2$
(from $F'$-preserving higher-order corrections to
$\mathcal{L}_{F'}$). It is important that this way of generating effective Yukawa
couplings is of topological origin and does not require a vacuum
alignment mechanism for $\phi_1$ and $\phi_2$: The VEVs
$\langle\phi_1\rangle$ and $\langle\phi_2\rangle$ can almost be chosen
arbitrarily, leading to the same qualitative result.

The global symmetry $F'$ is finally broken at some lower scale. This will produce additional small corrections $\epsilon''$ filling out the
zeros on the diagonals of the matrices in
(\ref{eq:F'preserving}). To avoid cosmological domain walls, $F'$ should, however, not be broken at a scale that is too low.

\section{Quark and Lepton Masses and Mixings}\label{sec:masses+mixings}
The actual Yukawa coupling matrices $Y^x_{ij}$ in (\ref{eq:4DLagrangian}) can
be viewed as resulting from adding to the Yukawa couplings
$\tilde{Y}_{ij}^x$ in (\ref{eq:quarktextures}) and
(\ref{eq:neutrinotextures}) the effect
of Wilson line breaking described in Sec.~\ref{sec:wilsonlines} as a
correction. For sufficiently small
$\epsilon'$ and $\epsilon''$, only $\tilde{Y}^d_{ij}$ and
$\tilde{Y}^\nu_{ij}$ will change after adding the corrections. The
matrices $\tilde{Y}_{ij}^u$ and $\tilde{Y}^R_{ij}$, on the other hand, will
practically remain unaltered. The leading order down quark and lepton
Yukawa couplings thus become
\begin{equation}\label{eq:fullquarktexture}
Y^d_{ij}=\left(
\begin{array}{ccc}
-1.4 \epsilon^5 & \epsilon^4 & -\epsilon^4\\
0 & 2\epsilon' & \epsilon''\\
\epsilon^7 & \epsilon & \epsilon
\end{array}
\right),
\end{equation}
\begin{equation}\label{eq:fullneutrinotextures}
Y^\nu_{ij}=\left(
\begin{array}{ccc}
0 & -\epsilon^4 & 0\\
-1.7\epsilon'' & -\epsilon^4 & 0.7\epsilon''\\
\epsilon'&\epsilon^4&\epsilon'
\end{array}
\right),\;
Y^R_{ij}=\epsilon^4\left(
\begin{array}{ccc}
1& 0 & 0.9\\
0 & 1.9\epsilon & 0\\
0.9 & 0 & 1
\end{array}
\right),
\end{equation}
while $Y^u_{ij}=\tilde{Y}^u_{ij}$ is as in
(\ref{eq:quarktextures}). Note that $Y^R_{ij}=\tilde{Y}^R_{ij}$ and we
have included in (\ref{eq:fullneutrinotextures}) an explicit form of
$Y^R_{ij}$ for a more detailed discussion of the PMNS observables
later. For certain values of $\epsilon'$ and $\epsilon''$, the order unity factors in
(\ref{eq:fullquarktexture}) and (\ref{eq:fullneutrinotextures}) allow
a viable fit to current neutrino data. After Wilson line breaking, the exact relation between the Yukawa couplings in
(\ref{eq:tilderelations}) is carried over to the matrix $Y^d_{ij}$.  The non-Abelian flavor symmetry therefore requires in  (\ref{eq:fullquarktexture}) that
\begin{equation}\label{eq:constraints}
 Y^d_{32}=Y^d_{33}.
\end{equation}
Note in (\ref{eq:fullneutrinotextures}) that the symmetry $G_1$ is
explicitly broken in the 1st row of $Y^d_{ij}$ and the 2nd column of $Y^\nu_{ij}$. As already mentioned, this can be arranged
by putting ${\bf 10}_1$ at a point on $S_1$ and ${\bf 1}_2$ at a point on $S_2$ where $G_1$ is not
conserved. In the following, we take $\epsilon'=\epsilon^3$ and
$\epsilon''=\epsilon^5$. Since $Y^u_{ij}$ is diagonal, the CKM angles
are entirely generated by $Y^d_{ij}$ which is on a lopsided form
\cite{lopsided}. As a result, the mass ratios of quarks and charged
leptons are given by
\begin{equation}\label{eq:quark+leptonmasses}
m_u:m_c:m_t=\epsilon^6:\epsilon^4:1,\quad
m_d:m_s:m_b=\epsilon^4:\epsilon^2:1,\quad
m_e:m_\mu:m_\tau=\epsilon^4:\epsilon^2:1,
\end{equation}
whereas the CKM angles become of the orders
\begin{equation}\label{eq:CKM}
V_{us}\sim\epsilon,\quad V_{cb}\sim\epsilon^2,\quad V_{ub}\sim\epsilon^3.
\end{equation}
The mass relations in (\ref{eq:quark+leptonmasses}) are a consequence
of minimal $SU(5)$. Extra Georgi-Jarlskog factors could be introduced
by extending the Higgs sector in a standard way \cite{Georgi:1979df}. In our example, we have a
moderate $\text{tan}\,\beta\sim 10$ and since the charged lepton
Yukawa coupling matrix is given by $Y^e_{ij}=Y^d_{ji}$, the model exhibits $b-\tau$
unification.

Light neutrino masses of the order $\sim 10^{-2}\,\text{eV}$ are
generated by the canonical type-I seesaw mechanism \cite{typeIseesaw} from the
effective neutrino mass matrix
\begin{equation}
M_\nu=-M_DM_R^{-1}M_D^T,
\end{equation}
where $(M_D)_{ij}=v\,\text{sin}\,\beta\,Y^\nu_{ij}$, with $v=174\,\text{GeV}$, and
$(M_R)_{ij}=M_{B-L} Y^R_{ij}$. Diagonalization of $M_\nu$ leads to a normal hierarchical neutrino mass spectrum
\begin{equation}
 m_1:m_2:m_3=\epsilon^2:\epsilon:1.\label{eq:neutrinomasses}
\end{equation}
The leptonic PMNS mixing matrix $U_\text{PMNS}=U_\ell^\dagger U_\nu$
is a product of the charged lepton mixing matrix $U_\ell$
(diagonalizing $Y^e{Y^e}^\dagger$) and the neutrino mixing matrix
$U_\nu$ (bringing $M_\nu$ on diagonal form). Both
contributions $U_\ell$ and $U_\nu$ are important in $U_\text{PMNS}$, since the
left-handed charged leptons as well as the neutrinos exhibit large mixings. Numerically, for our fit
in (\ref{eq:fullquarktexture}) and (\ref{eq:fullneutrinotextures}), the solar,
atmospheric, and reactor mixing angle, respectively take the values\footnote{The uncertainty in $Y^R_{13}=0.9\cdot\epsilon^4$ leads to an uncertainty in $\theta_{12}$ of $\sim 3^\circ$ around the central value of $37^\circ$.}
\begin{equation}
\theta_{12}\approx 36^\circ,\quad\theta_{23}\approx
53^\circ,\quad\theta_{13}\approx 7^\circ.
\end{equation}
The Yukawa coupling matrices in (\ref{eq:fullquarktexture})
  and (\ref{eq:fullneutrinotextures}) therefore yield neutrino masses and
  mixings in agreement with current data at the $2\sigma$ level. The PMNS mixing angles originate from the relations\footnote{Related sum rules in
  connection with nonzero CP-phases have also been discussed in \cite{Niehage:2008sg}.}
\begin{equation}\label{eq:sumrules}
\theta_{12}\approx\pi/4-2\epsilon^2,\quad
\theta_{23}\approx\pi/4+\epsilon/2,\quad\theta_{13}\approx\epsilon/2,
\end{equation}
where we have kept only the dominant contribution to
$\theta_{13}$. From (\ref{eq:sumrules}), we see that the solar angle
follows a modified quark-lepton complementarity relation
\cite{qlc} and that the reactor angle is about half the Cabibbo
angle, which could be tested in next-generation experiments
\cite{Huber:2006vr}. In (\ref{eq:sumrules}), it is important that in the
expansion of the atmospheric angle $\theta_{23}$ the non-Abelian
flavor symmetry predicts the zeroth order term to be exactly $\pi/4$. The deviation from maximal atmospheric mixing
introduced by the correction $\sim\epsilon/2$ drives $\theta_{23}$ to
a larger value, which may be testable in future neutrino oscillation
experiments such as NO$\nu$A, T2K, or a neutrino factory \cite{Ayres:2004js}.

So far, we have been considering only real Yukawa couplings. An excellent fit of $\theta_{12}$ and $\theta_{23}$ at the
  $1\sigma$ level while keeping $\theta_{13}$ small, however, can be achieved by including nonzero CP phases. In Fig.~1, we have randomly
  varied (scattered) the phases of the Yukawa couplings in (\ref{eq:fullquarktexture})
  and (\ref{eq:fullneutrinotextures}), while respecting the constraint in
  (\ref{eq:constraints}) imposed by the non-Abelian flavor symmetry. In this scattering, we have kept the moduli of
  the Yukawa couplings fixed and required that the lepton mass ratios
  in (\ref{eq:quark+leptonmasses}) and (\ref{eq:neutrinomasses}) are
  satisfied within relative factors of at most
  $1.5$. For each point in Fig.~1, $\theta_{12}$ and $\theta_{23}$ are within the current $1\sigma$
  bounds and have $\theta_{13}\leq 5^\circ$,
  simultaneously. While $\theta_{12}$ and $\theta_{23}$ are
  distributed over the whole $1\sigma$ intervals, very small values
  $\theta_{13}<4^\circ$ are seldom in this scatter. Fig.~1 shows also the accompanying low energy  Dirac and Majorana CP phases $\delta$ and
  $\phi_{1,2}$, which can be large (middle and right panel). Note that, interestingly, the Majorana
  phases show for our matrices a preference for a crude correlation $\phi_1\sim\phi_2$. (The plots in Fig.~1 have been produced based on work done in \cite{Ruckl}.)
\begin{figure}[t]
\centering
\includegraphics[width=0.32\textwidth]{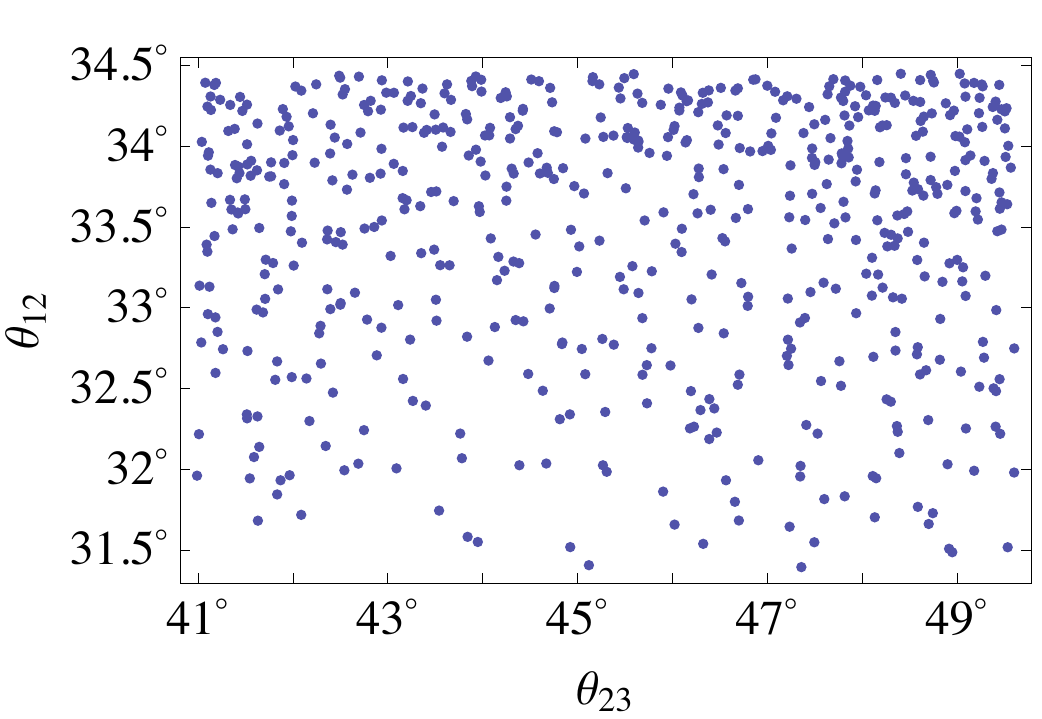}
\includegraphics[width=0.32\textwidth]{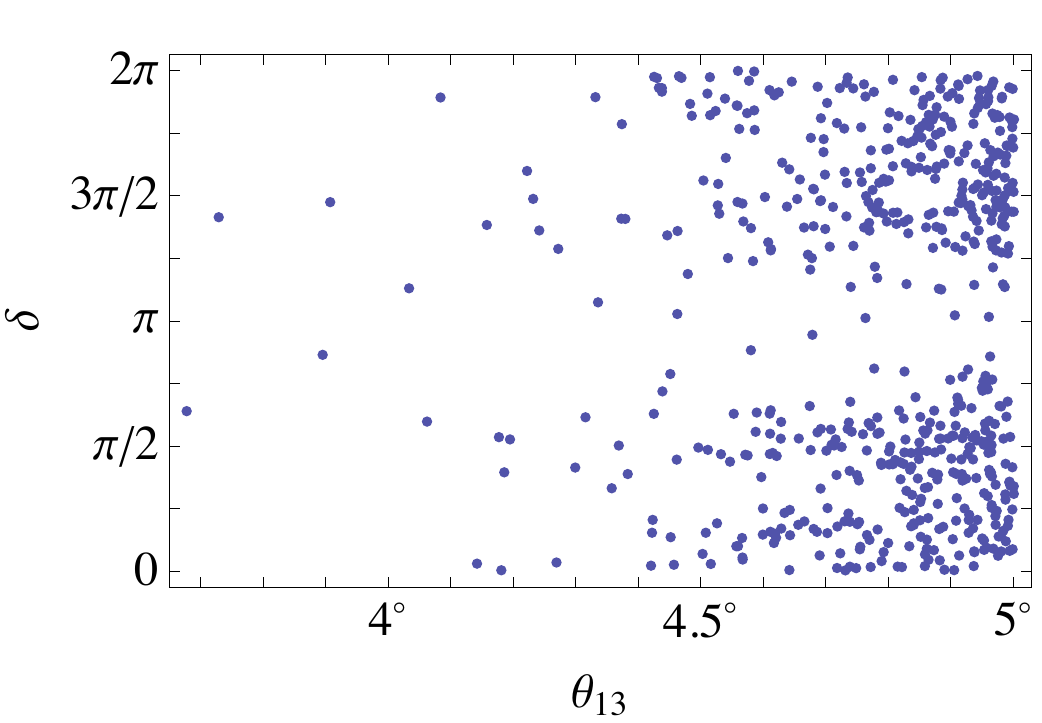}
\includegraphics[width=0.32\textwidth]{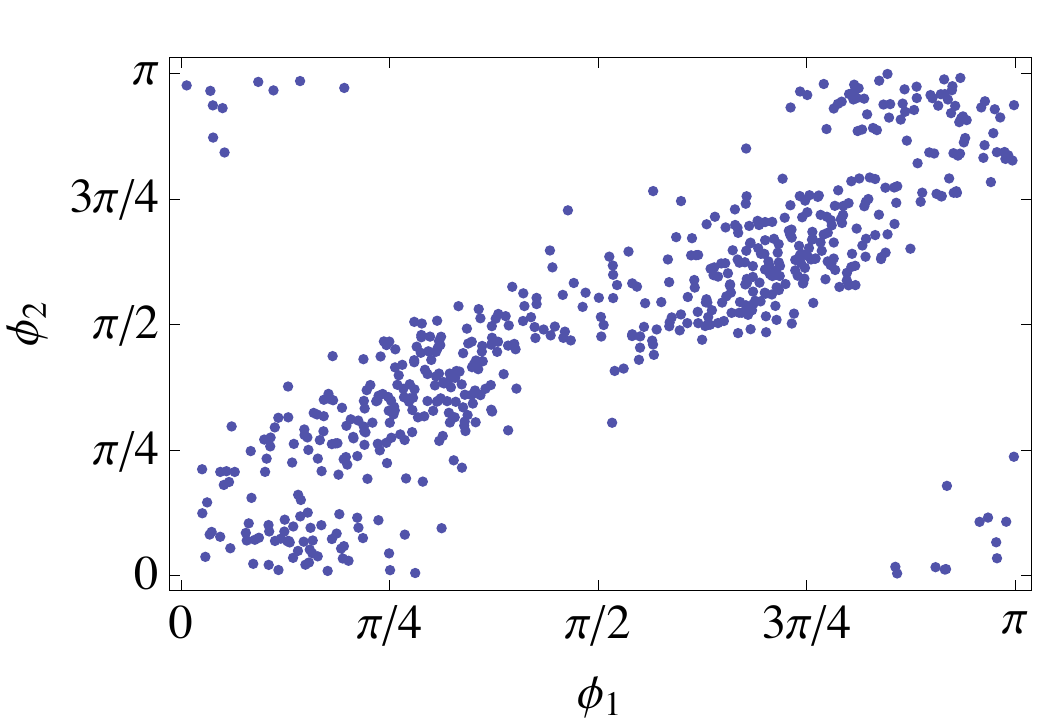}
\caption{PMNS mixing angles and CP phases in the model close to the
  best fit values. Shown are the observables obtained by randomly varying
  the phases of the Yukawa couplings in (\ref{eq:fullquarktexture})
  and (\ref{eq:fullneutrinotextures}) while respecting the constraint in
  (\ref{eq:constraints}). Both the solar and the
  atmospheric mixing angles are within the current $1\sigma$ ranges
  (see \cite{Schwetz:2008er}) and the reactor angle is very small,
  i.e.~$\theta_{13}\leq 5^\circ$. The Dirac phase $\delta$ (middle panel) and the
  Majorana phases $\phi_1$ and $\phi_2$ (right panel) can be
  large. Interestingly, $\phi_1$ and $\phi_2$ seem to exhibit a rough
  correlation $\phi_1\simeq\phi_2$.}
\end{figure}

The flavor symmetry group $G_F$ is broken at high energies such as the
GUT scale. Under renormalization group running from the
Planck scale down to low energies, however, the Cabibbo angle
$\theta_\text{C}$ is essentially stable and $V_{cb}\sim\epsilon^2$
changes by a factor less than 2 \cite{Arason:1991hu}. Also,
since we have a normal hierarchical light neutrino mass spectrum, renormalization
group effects have practically no influence on the mass ratios in
(\ref{eq:neutrinomasses}) and change the PMNS mixing angles only
by $\ll 1^\circ$ (for a discussion and further references see
\cite{Plentinger:2007px}). Given the precision of our model, we
therefore neglect the impact of renormalization group effects on our
results.

Proton decay via $d=5$ and $d=6$ operators as well as doublet-triplet splitting depend crucially on the geography of the matter fields and on the way
in which $SU(5)$ is broken in the extra dimensions (see, e.g., \cite{Kawamura:2000ev,protondecay}). A detailed study
of these issues is therefore beyond the scope of this paper and has to be
addressed elsewhere. Moreover, since $G_F$ is gauged, one may
cancel anomalies by adding Chern-Simons terms in the bulk.
 
\section{Summary and Conclusions}\label{sec:conclusions}
In this paper, we have presented a supersymmetric $SU(5)$ GUT toy model
with a discrete non-Abelian flavor symmetry that is broken by Wilson lines. The
model is formulated in 4+3 dimensions compactified on a manifold
$S^3/Z_n$. Wilson line breaking of the non-Abelian flavor symmetry is
topological and has the advantage that one can have both exact predictions
for fermion mixing angles and predictions for the
observed fermion mass hierarchies without a vacuum
alignment mechanism.

In conjunction with the Froggatt-Nielsen mechanism, the model produces the
hierarchical pattern of quark masses and CKM angles. The CKM matrix is
entirely generated by the down quark mass matrix which is on a
lopsided form, i.e.~large atmospheric mixing comes mainly from the
charged leptons. In the lepton sector, we obtain the hierarchical
charged lepton mass spectrum and normal hierarchical neutrino masses that become small via the type-I seesaw mechanism with three heavy right-handed
neutrinos. The PMNS angles are in excellent agreement with current
data at $1\sigma$, exhibiting values that could be tested in future neutrino
oscillation experiments. The solar angle satisfies the quark-lepton complementarity-type relation
$\theta_{12}\approx\pi/4-2\theta_\text{C}^2$ while the reactor
angle is about half the Cabibbo angle. We have shown that the
inclusion of nonzero phases of the Yukawa couplings allows large
low-energy Dirac and Majorana CP phases. In particular, we have found
that the two Majorana phases exhibit a rough correlation $\phi_1\sim \phi_2$. After Wilson line breaking, the non-Abelian flavor symmetry predicts a maximal atmospheric mixing angle which is driven to
larger values by a correction $\sim \theta_\text{C}/2$. The simultaneous prediction of (i)
nearly maximal atmospheric mixing from the non-Abelian flavor symmetry
and (ii) the strict mass hierarchy between the 2nd and 3rd generation of down-type
charged fermions is of topological origin: We have non-trivial flavon
representations but they can take almost arbitrary VEVs giving
practically the same result, i.e.~there is no need for a vacuum alignment mechanism.

We have focussed on a specific non-Abelian example flavor symmetry
with two-dimensional representations, but it would be desirable to apply flavor symmetry breaking by Wilson
lines also to other symmetries such as $A_4,T',$ or
$\Delta(3n^2)$, admitting three-dimensional representations. In
this way, one could try to arrive at additional exact predictions for the
PMNS mixing angles. Finally, it would also
be interesting to see how Wilson line flavor symmetry breaking can be
formulated for other GUTs such as $SO(10)$ or $E_6$. 

\section*{Acknowledgments}
This work was supported by the Federal Ministry of Education and Research (BMBF) under contract number 05HT6WWA.

\end{document}